\documentclass[conference]{IEEEtran}
\IEEEoverridecommandlockouts
\usepackage{cite}
\usepackage{amsmath,amssymb,amsfonts}
\usepackage{algorithmic}
\usepackage{graphicx}
\usepackage{textcomp}
\usepackage{xcolor}
\usepackage{scrextend}
\usepackage[hidelinks]{hyperref}
\def\BibTeX{{\rm B\kern-.05em{\sc i\kern-.025em b}\kern-.08em
    T\kern-.1667em\lower.7ex\hbox{E}\kern-.125emX}}
\usepackage{todonotes}
\usepackage[nolist]{acronym}
\usepackage{url}
\usepackage[strings]{underscore}
\usepackage{enumitem}\usepackage{selinput}
\usepackage{colortbl}
\usepackage[breakable, hooks,most]{tcolorbox}
\usepackage{makecell}

\usepackage{booktabs}

\usepackage{rotating}
\usepackage{longtable}
\usepackage{array}
\usepackage[longtable]{multirow}
\usepackage{pdflscape}\usepackage{tabularray}
\usepackage{makecell}
\usepackage{setspace}
\usepackage{pifont}

\usepackage{color}
\usepackage{enumitem}
\usepackage{colortbl}
\usepackage[breakable, hooks]{tcolorbox}

\begin{document}
\bstctlcite{IEEEexample:BSTcontrol}

\begin{acronym}
    \acro{AI}{Artificial Intelligence}
    \acro{ML}{Machine Learning}
    \acro{DL}{Deep Learning}
    \acro{CD4ML}{Continuous Delivery for Machine Learning}
    \acro{Dev}{Development}
    \acro{GLR}{Grey Literature Review}
    \acro{TFX}{Tensorflow Extended}
    \acro{MLR}{Multivocal Literature Review}
    \acro{MLOps}{Machine Learning Operations}
    \acro{SMS}{Systematic Mapping Study}
    \acro{SE}{Software Engineering}
    \acro{QA}{Quality Assurance}
    \acro{VM}{Virtual Machine}
    \acro{NFS}{Network File System}
    \acro{MSC}{Micro-sized Comp.[$\geq$10 employees]}
    \acro{SC}{Small Comp. [employees$\leq$49]}
    \acro{MC}{Medium Comp. [50$\leq$employees$\leq$249]}
    \acro{LC}{Large Comp. [employees$\geq$250]}
    \acro{EC}{Enterprise-level Comp. [LC + 1bn € revenue]}
    \acro{PLC}{Programmable Logic Controller}
    \acro{RAM}{Random-Access Memory}
    \acro{PLS-SEM}{Partial Least Squares Structural Equation Modeling} 
    \acro{RAPL}{Running Average Power Limit}
\end{acronym}

\newcommand{\customnewline}[1][2pt]{\vspace*{#1}}
\newcommand{\columTextSize}{\footnotesize}

\definecolor{mylabelcolor}{RGB}{120, 27, 124}
\definecolor{mylabelcolorExamples}{RGB}{120, 27, 124}
\definecolor{fillColor}{RGB}{240,240,240} 
\definecolor{frameColor}{RGB}{138,147,156} 

\definecolor{mynewcolor}{HTML}{1E90FF} 
\newcommand{\change}[1]{\textcolor{mynewcolor}{#1}}

\definecolor{logs}{HTML}{869FB2}
\definecolor{traces}{HTML}{B2869F}
\definecolor{metrics}{HTML}{9FB286}

\newtcolorbox{mycolorbox}[2][]{
  enhanced, 
  attach boxed title to top left={
        xshift=-0.1cm,
        yshift=-\tcboxedtitleheight/2},  
    top=4mm,
  breakable,
  boxsep=-0.1pt,
  outer arc=2pt,
  colback=fillColor,
  colframe=frameColor,
  title={\textcolor{black}{\textbf{#2}}},
  colbacktitle=white,
  coltitle=black,
  grow to left by=0mm,
  left*=1mm,
  grow to right by=0mm,
  right*=2mm,
  before skip=2pt, 
  after skip=2pt,
  #1
}

\newtcolorbox{mycolorboxRQ}[1][]{
  enhanced,
  breakable,
  boxsep=-0.1pt,
  outer arc=2pt,
  colback=white,
  colframe=frameColor,
  colbacktitle=white,
  coltitle=black,
  grow to left by=0mm,
  left*=1mm,
  grow to right by=0mm,
  right*=2mm,
  before skip=4pt, 
  after skip=4pt,
  #1
}

\makeatletter
\newcommand{\labelInterviewee}[2]{%
   \protected@write \@auxout {}{\string \newlabel {#1}{{\textcolor{mylabelcolor}{#2}}{\thepage}{\textcolor{mylabelcolor}{\textbf{#2}}}{#1}{}} }%
   \hypertarget{#1}{\textcolor{mylabelcolor}{\textbf{#2}}}
}
\makeatother

\makeatletter
\newcommand{\labelExamples}[2]{%
   \protected@write \@auxout {}{\string \newlabel {#1}{{\textcolor{mylabelcolorExamples}{#2}}{\thepage}{\textcolor{mylabelcolorExamples}{\textbf{#2}}}{#1}{}} }%
   \hypertarget{#1}{\textcolor{mylabelcolorExamples}{\textbf{#2}}}
}
\makeatother

\title{
How Industry Tackles Anomalies during Runtime: Approaches and Key Monitoring Parameters

\thanks{This work was supported by the Austrian Research Promotion Agency (FFG) in the frame of the ConTest project [888127] and the SmartQuality (SmartDelta) project [890417]. We thank the study participants, including Vaadin, Akkodis, Turkcell Technology, Hoxhunt, and all other anonymized partners, for their valuable insights.}}

\author{\IEEEauthorblockN{Monika Steidl\IEEEauthorrefmark{1}, Benedikt Dornauer\IEEEauthorrefmark{2}, Michael Felderer\IEEEauthorrefmark{3},\\ Rudolf Ramler\IEEEauthorrefmark{4}, Mircea-Cristian Racasan\IEEEauthorrefmark{5}, Marko Gattringer\IEEEauthorrefmark{6}}
\IEEEauthorblockA{\IEEEauthorrefmark{1}\IEEEauthorrefmark{2}\IEEEauthorrefmark{3}\textit{University of Innsbruck}, Innsbruck, Austria}
\IEEEauthorblockA{\IEEEauthorrefmark{4}\textit{Software Competence Center Hagenberg GmbH}, Hagenberg, Austria}
\IEEEauthorblockA{\IEEEauthorrefmark{3}\textit{German Aerospace Center (DLR), Institute of Software Technology}, Cologne, Germany}
\IEEEauthorblockA{\IEEEauthorrefmark{5}\textit{c.c.com Moser GmbH}, Grambach, Austria}
\IEEEauthorblockA{\IEEEauthorrefmark{6}\textit{Gepardec IT Services GmbH}, Linz, Austria}
\IEEEauthorblockA{\IEEEauthorrefmark{2}\IEEEauthorrefmark{3}\textit{University of Cologne}, Cologne, Germany}
\IEEEauthorblockA{ORCID: \IEEEauthorrefmark{1}0000-0002-3410-7637, \IEEEauthorrefmark{2}0000-0002-7713-4686,\IEEEauthorrefmark{3}0000-0003-3818-4442,\\ \IEEEauthorrefmark{4}0000-0001-9903-6107,\IEEEauthorrefmark{5}0009-0008-7938-3126,\IEEEauthorrefmark{6}0000-0003-1659-3624}
}
\maketitle

\begin{abstract}
Deviations from expected behavior during runtime, known as anomalies, have become more common due to the systems' complexity, especially for microservices. Consequently, analyzing runtime monitoring data, such as logs, traces for microservices, and metrics, is challenging due to the large volume of data collected. Developing effective rules or AI algorithms requires a deep understanding of this data to reliably detect unforeseen anomalies. 
This paper seeks to comprehend anomalies and current anomaly detection approaches across diverse industrial sectors. Additionally, it aims to pinpoint the parameters necessary for identifying anomalies via runtime monitoring data.
Therefore, we conducted semi-structured interviews with fifteen industry participants who rely on anomaly detection during runtime. Additionally, to supplement information from the interviews, we performed a literature review focusing on anomaly detection approaches applied to industrial real-life datasets.
Our paper (1) demonstrates the diversity of interpretations and examples of software anomalies during runtime and (2) explores the reasons behind choosing rule-based approaches in the industry over self-developed AI approaches. AI-based approaches have become prominent in published industry-related papers in the last three years. Furthermore, we (3) identified key monitoring parameters collected during runtime (logs, traces, and metrics) that assist practitioners in detecting anomalies during runtime without introducing bias in their anomaly detection approach due to inconclusive parameters.
\end{abstract}


\section{Introduction}
\label{sec:Introduction}

In the late 1970s and early 1980s, system administrators manually inspected printed audit logs, which often piled up to four to five feet by week's end, to search for suspicious behavior. This might mark the start of software anomaly detection\cite{1012428}. Nowadays, frequent software changes have become the industry norm for fixing bugs, improving performance, and enhancing user satisfaction with new features \cite{sohist}. Due to these extensive changes, classical test suites frequently cannot prevent all potential \textit{deviations from expected behavior}\cite{ieeeDefAnomaly1990} at runtime, often described as anomalies. Beyond software issues, system reliance on hardware may also result in system failures or performance declines from processing queue saturation \cite{Islam2021}. Mariani et al. \cite{Mariani2020} even stated that runtime anomalies are becoming the norm rather than the exception in various systems (e.g., ultra-large systems, system of systems, or cloud systems). Therefore, the industry requires effective strategies to preserve the integrity and functionality of its software systems even during runtime. 

These effective strategies must handle volatile anomalies with different root causes and varying observable behavior. Currently, there is no consensus on a classical notion of software anomalies during runtime \cite{Samariya2023}. This makes it hard to consistently identify and communicate anomalies, complicating efforts to ensure software quality, prioritize the handling of specific anomalies, and manage risks based on severity and impact. Thus, it is crucial to understand how the industry interprets and characterizes these anomalies. In addition, gaining more insights into potential anomalies by extending related work \cite{Zhou2021,Steidl2022,Silva2022,Zhang2023RobustFailureDiagnosis,Xie2023,Xie2023a,Wang2018,Lee2023,Zhang2023,Li2021,Liu2021} with further experienced real-life anomalies, allows to enhance awareness of further potential anomalies:

\begin{mycolorboxRQ}
\textbf{RQ1:} What are the prevailing interpretations, characteristics, and examples of anomalies within the industry? \end{mycolorboxRQ}

Anomaly detection during runtime involves continuously monitoring and analyzing a system's operations to identify the underlying reasons, also known as root cause, for anomalies in time or before the system is affected. Thus, developers frequently find themselves manually examining a huge volume of runtime monitoring data, such as logs, traces (for microservices), or metrics that exhibit big data characteristics \cite{Soldani2022}. Not only is analyzing this runtime monitoring data labor-intensive due to a large amount of available data and challenging due to the system's complexity (e.g., constant changes in traffic, scaling requirements, complex infrastructures), but behavior that appears anomalous may not always signify an actual anomaly \cite{Steidl2022,Li2021,Li2022}. To ease this cumbersome work, semi-automated approaches exist to assist developers and operators in detecting software anomalies during runtime. These approaches rely on self-defined rules and thresholds and \ac{AI} algorithms \cite{Islam2021,Liu2020,Jin2020,Lee2023} to detect anomalies and their respective root cause. We examined companies' choices and rationales to understand why they opted for \textbf{rule-based} or \textbf{\ac{AI}-based} approaches:

\begin{mycolorboxRQ}
\textbf{RQ2:} What factors influence the selection of anomaly detection approaches in industrial settings?\end{mycolorboxRQ}

Both approaches heavily rely on a high-quality dataset and domain-specific knowledge \cite{Ikram2023,Chen2023} where a consensus on which runtime monitoring parameter indicates an anomaly is missing \cite{DeSilva2022}.
In related work, only a few anomaly detection algorithms explicitly indicate which parameters are used as input dataset \cite{Lee2023,Li2021,Mantyla2023,Zhang2023RobustFailureDiagnosis}. Thus, it is difficult to replicate published performance measurements (e.g., accuracy, precision, and recall) of open-source AI-based anomaly detection methods, indicating their high dependency on optimized hyperparameters and a comprehensive monitoring dataset that includes all essential parameters without introducing biases that could skew the results \cite{Steidl2022}. 

Parameters are extracted from runtime monitoring data, such as logs, traces (for microservices), and metrics. Logs consist of predefined semi-structured emitted messages with natural language, traces are microservice-specific data types representing the end-to-end execution of a single request, and metrics consist of numeric time-series performance data. By knowing which parameters are commonly considered, anomaly detection approaches could better suit industry needs and improve their effectiveness, leading to more robust and targeted rules or training datasets for AI-based approaches \cite{Ma2020}. The importance of recognizing essential parameters derived from runtime monitoring data is often overlooked. For instance, related work has focused on a single type of runtime monitoring data for detecting anomalies in microservices \cite{Jia2017,Jia2017-LogSed, Liu2020,Nedelkoski2019,Jin2020,Nedelkoski2019a,Wang2018}, and focusing on the combination of these gain increased attention \cite{Zhang2023RobustFailureDiagnosis,Lee2023,Liu2021}. Understanding the relationships between these parameters (e.g., an increased response rate increases CPU usage) and the dependencies among microservices is essential for avoiding false positives. Stable relationships may indicate the absence of anomalies. This understanding allows removing unnecessary parameters to optimize storage and computational resources without relying on extensive manual work \cite{Steidl2022}. Thus, our RQ is: 
\begin{mycolorboxRQ}
\textbf{RQ3:} Which runtime monitoring data is used to identify anomalies by industry?
\end{mycolorboxRQ}

The remainder of the paper is structured as follows: Section \ref{sec:Methodology} describes the study design of the literature review and semi-structured interviews. Section \ref{sec:Results} addresses and discusses the RQ 1-3. Afterward, Section \ref{sec:threatsToValidity} discusses potential threats, followed by a summary and future work in Section \ref{sec:conclusion}.

\section{Study design}
\label{sec:Methodology}
We applied two research methods to answer our defined research questions. Firstly, we extended an existing (A) literature review with additional anomaly detection approaches for microservices related to industry use cases and conducted (B) semi-structured interviews. We followed the methods regarding the Empirical Standards for Software Engineering Research v2.0 \cite{Ralph2020}. For detailed information, please refer to the replication package in \cite{Zenodo}. Due to confidentiality and company regulations, we cannot disclose the full recorded interviews and transcripts.

\subsection{Extended Literature Review}
\label{sec:systematicReview}
In February 2022, Soldani and Brogi \cite{Soldani2022} published the first structured overview of current research regarding anomaly detection during runtime, root cause analysis, and required algorithms, specifically addressing the context of microservices. We also specifically included microservices because this additional monitoring data type, traces, is highly relevant for anomaly detection due to its popularity in the last years \cite{Li2022}. However, their stated methodology was a survey without systematically collecting literature based on a search string, which we aim to extend with our approach.

Thus, we extended their list of anomaly detection approaches with the search term "\textit{anomaly detection runtime monitoring microservices}" where the search was executed via Google Scholar. We started in June 2023 to gain further insights into current research gaps. To consider the latest research advances, we again proceeded with the same literature study from December 2023 to January 2024, where we verified previously obtained literature and added new ones published within this timeframe. We included peer-reviewed anomaly detection algorithms for microservices published after 2015. 
To assess if our search string includes relevant keywords to overlap with the papers cited in Soldani and Brogi \cite{Soldani2022}, we compared our findings with their papers, where we were able to identify exactly 50\%. However, we discovered 30 additional papers published before February 2022 that Soldani and Brogi did not include. In total, we found 92 papers focusing on anomaly detection for microservices. Of these, 36 papers were considered for further argumentation in this paper due to their evaluation based on real-life datasets and industry relevance. We excluded works that evaluated their approach with benchmark systems due to their artificial setting and injected anomalies. We provide details about the selection process and the included and excluded papers in the replication package \cite{Zenodo}.

To minimize selection bias, the second author executed a blind review of the included and excluded papers and categorized industry-relevant papers, where we achieved an inter-rater reliability of 94.7\%. To resolve discrepancies, we discussed the decisions until we reached an agreement.

\subsection{Semi-structured Interviews in Various Domains}
\label{sec:semiStructuredInterviews}
We followed the guidelines by Runeson and Höst \cite{Runeson2009} for our semi-structured interviews and relied on recommendations for software engineering interviews by Hove and Anda \cite{Hove2005}. 

We explicitly focused on interview participants from various company sizes (balancing between small and medium enterprises and large companies), monolithic and microservice architecture (for tracing information), domains (holistic overview), and experiences to gain domain unspecific insights (heterogeneous sample), as outlined in Table \ref{tab:interviewees}. We used purposive sampling to contact companies from the authors' network that rely on anomaly detection during runtime. The goal was to gain insights into experienced anomalies, aiming to show that anomalies and their effects are highly volatile and unpredictable. Furthermore, we observed factors influencing the selection of anomaly detection approaches and key monitoring parameters. Thus, the goal is not to provide statistical inference but to gain and discuss qualitative insights. Therefore, we asked for skilled experts who either work as developers, data, or DevOps Engineers or ensure software quality. These interview participants need to be involved in the manual or automated identification of anomalies happening during runtime based on logs, traces (for microservices), or metrics. Furthermore, along with our request, we attached our interview guidelines. 
\begin{table}[htbp]
\centering  
\normalsize 
\vspace{-0.3cm}
\resizebox{9.0cm}{!}{
    \begin{tabular}{m{0.15cm}m{0.10cm} m{0.2cm} m{0.46cm} m{3cm} m{4.5cm} m{0.8cm} m{0.3cm}m{0.3cm}m{0.3cm}} 
    \hline
    \rowcolor{fillColor} & \textbf{ID} & \textbf{} & \textbf{Size}\footref{1sttablefoot} &\textbf{Specific Domain} & \textbf{Experience} & \textbf{Years} & \textbf{RU}\footref{1sttablefoot}& \textbf{AI}\footref{1sttablefoot} & \textbf{MS}\footref{1sttablefoot}\\
    \hline
     \parbox[t]{2pt}{\multirow{6}{*}{\rotatebox[origin=c]{90}{\textbf{Finance}}}}  & \labelInterviewee{B}{B} & \vspace*{2pt}\frame{\includegraphics[width=0.4cm]{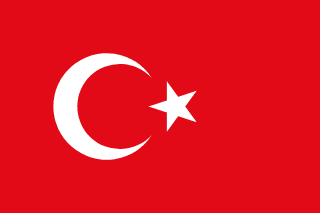}} & \acs{LC} & Finance Payment B2B Provider &  Linux-Sysadmin and R\&D manager & 13 \& 2 & \checkmark  &  $\sim$  \\ 
     \cline{2-10}
    & \labelInterviewee{F}{F} &  \vspace*{2pt}\frame{\includegraphics[width=0.4cm]{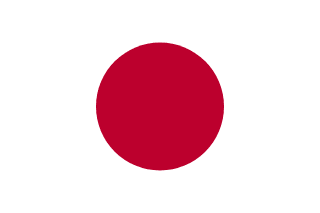}} & \acs{LC} & Global Financial Services Group &  Data Engineer and Scientist for Log Anomalies & 14 & \checkmark &  \\ 
    \cline{2-10}
    &  \labelInterviewee{M}{M} & \multirow{3}{*}{\frame{\includegraphics[width=0.4cm]{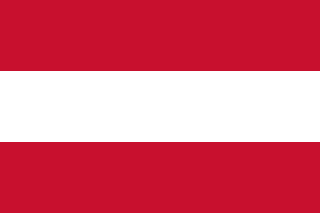}}}  & \multirow{3}{*}{\acs{MC}} & \multirow{3}{*}{\parbox{4.5cm}{\raggedright Financial Service\\ Provider}} & Product Owner for Software Quality, before Load Tester& 18  & \checkmark &  $\sim$ & \checkmark\\ 
    &  \labelInterviewee{N}{N} & & &  & Load Tester and Release Automation Engineer & 5 & \checkmark &  $\sim$ & \checkmark \\ 
    \hline 
     \parbox[t]{2mm}{\multirow{6}{*}{\rotatebox[origin=c]{90}{\textbf{Web}}}}  &  \labelInterviewee{A}{A} & \vspace*{2pt}\frame{ \includegraphics[width=0.4cm]{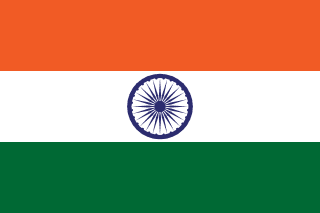}} & \acs{SC} & End-user Web Application  & DevOps Engineer & 20 & \checkmark &  \\ 
     \cline{2-10}
    & \labelInterviewee{G}{G} & \multirow{ 3}{*}{\frame{\includegraphics[width=0.4cm]{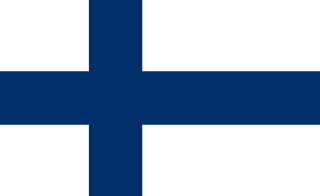}}} & \multirow{ 3}{*}{\acs{MC}} & \multirow{3}{*}{\parbox{4.5cm}{\raggedright Human Risk \\Management \\Platform}} & 
Junior Security Engineer skilled in \ac{AI} anomaly detection & 3 & \checkmark & \checkmark & \checkmark\\ 
    & \labelInterviewee{J}{J} & & & & Site Reliability Engineer for Cloud Applications & 3 & \checkmark & \checkmark & \checkmark\\ 
    \cline{2-10}
    & \labelInterviewee{L}{L} &  \vspace*{2pt}\frame{\includegraphics[width=0.4cm]{images/flags/fi.png}}  & \acs{MC} & Web Application Framework Provider & Product Developer & 14 & \checkmark & & \checkmark\\ 
    \hline 
   \parbox[t]{2pt}{\multirow{5}{*}{\rotatebox[origin=c]{90}{\textbf{Hardware}}}}  & \labelInterviewee{D}{D} & \vspace*{2pt} \frame{\includegraphics[width=0.4cm]{images/flags/at.png}}  & \acs{SC} & Traffic Analysis System Provider& Head of Software Engineering & 16 & \checkmark & & \checkmark \\ 
    \cline{2-10}
    & \labelInterviewee{I}{I} & \vspace*{2pt} \frame{\includegraphics[width=0.4cm]{images/flags/at.png}} & \acs{LC} & Manufacturing of Machinery & Team Lead with focus on quality assurance for \acf{PLC} & 10+ & \checkmark & 
    \\ 
    \cline{2-10}
    & \labelInterviewee{O}{O} & \vspace*{2pt} \frame{\includegraphics[width=0.4cm]{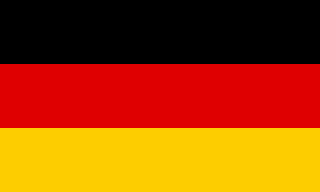}}  & \acs{LC} &  Global Digital Engineering Company& Embedded Software Team Lead & 9 & \checkmark &  \\ 
    \hline 
     \parbox[t]{2mm}{\multirow{6}{*}{\rotatebox[origin=c]{90}{\textbf{Other}}}}   & \labelInterviewee{C}{C} & \vspace*{2pt} \frame{\includegraphics[width=0.4cm]{images/flags/at.png}} & \acs{MC} & Java/JEE Solutions and Cloud Tech. & Cloud Architect & 5 & \checkmark & $\sim$ \\ 
     \cline{2-10}
    & \labelInterviewee{E}{E}& \vspace*{2pt} \frame{\includegraphics[width=0.4cm]{images/flags/tr.png}}  & \acs{LC}  & Tele. Company and Network Provider & Senior DevOps Engineer & 3 & \checkmark & $\sim$ & \checkmark \\ 
    \cline{2-10}
    &\labelInterviewee{H}{H} & \multirow{ 3}{*}{\frame{\includegraphics[width=0.4cm]{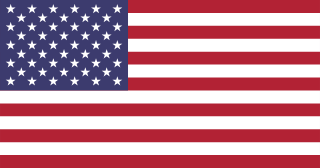}}} & \multirow{ 3}{*}{\acs{LC}} & \multirow{3}{*}{\parbox{4.5cm}{\raggedright Software \\ Observability \\ Platform}}& System and DevOps Engineer & 5 & \checkmark & \checkmark & \checkmark \\ 
    & \labelInterviewee{K}{K} &  &  & & Senior Data Scientist for univariate time series & 4 & \checkmark & \checkmark & \checkmark \\ 
    \hline
    \end{tabular}
    }
    \vspace{0.1cm}
\caption{List of interview participants with IDs assigned alphabetically and listed chronologically.} 
\label{tab:interviewees}
\vspace{-0.45cm}
\end{table}

\footnotetext[1]{\label{1sttablefoot}Company Size based on OECD\cite{OECD2017}: \acf{SC}, \acf{MC}, \acf{LC}. Rule-based (RU) approach. AI-based (AI) approach [$\sim$ = commercially available AI algorithms]. Microservice (MS).}
In total, we interviewed 15 participants from 12 different companies. Based on their products and services, the participants can be assigned to four categories: Finance software (software solutions for the financial sector), Web service (web-based service for end users), Technology hardware (hardware-based software or embedded systems), and Others (ranging from services to develop, integrate, and maintain custom Java/JEE solutions and cloud technologies, telecommunication, software observability), illustrating the variety of domains. 

Before the semi-structured interviews, we pilot-tested the guidelines with two individuals. This process allowed us to identify potential misunderstandings or comprehension difficulties and make necessary adjustments to the interview guidelines. The first two authors were present during the interviews, one responsible for guiding the interview and the other for asking additional clarification questions while taking notes. Every interview lasted at least 30 and up to 60 minutes and was conducted between October 2023 and February 2024.

\subsection{Data Extraction}
For both \hyperref[RQ1]{RQ1} and \hyperref[RQ2]{RQ2}, we applied inductive coding by extracting relevant information from the transcribed interviews, done independently, and afterward applied color schemes.


In particular, for \hyperref[RQ1]{RQ1}, we were interested in seeing how the collected interview data compares to available standards based on \cite{ieeeAnomalyStd2010} and if there exist common interpretations and characteristics among the industry. The main identified codes are \textit{argue that it is difficult to define anomaly, deviation of expectation, outliers of specific trends, synonym, negative effect, unforeseen or not reproducible, examples.}

For \hyperref[RQ2]{RQ2}, we were interested in the type(s) of anomaly detection approaches and their specific evaluation. Therefore, the main identified codes are \textit{rule-based, AI-based, both} and \textit{advantage, disadvantage}.

For \hyperref[RQ3]{RQ3}, we executed deductive coding of the collected papers from the literature review and transcribed interviews because we wanted to base the identified data on preconceived categories that are based on theory and existing knowledge. Soldani and Brogi \cite{Soldani2022} categorized the input for anomaly detection approaches into logs, traces, and metrics. Therefore, the main identified codes for logs are \textit{static part, error/warnings, others}. For traces, the codes are \textit{HTTP status code, depth of microservice invocation path, similarity between control flow graph, response time}, and for metrics, the codes are \textit{queues, CPU, memory, network traffic, disk, energy consumption}. In RQ3, we present the overlap between literature and interviews, also considering the differences in stated parameters between literature and interview findings.

For all three RQs, if discrepancies arose, the authors resolved them through discussion where respective codes were constantly reviewed and adapted to allow an accurate data representation. Furthermore, we evaluated the coded results by consolidating with the interview participants afterward. We provide all interview materials, including questions and coding procedures, in the replication package \cite{Zenodo}.  

\section{Findings and Discussion}
\label{sec:Results}
In the following, we illustrate and discuss interpretations, characteristics, and examples of anomalies by industry (\hyperref[RQ1]{RQ1}) and further elaborate existing anomaly detection approaches in the industry with their advantages and shortcomings (\hyperref[RQ2]{RQ2}). These approaches rely on key monitoring parameters to gain insights into deviant behavior. Therefore, (\hyperref[RQ3]{RQ3}) gathers industry-relevant parameters.

\subsection*{RQ1: Interpretations, Characteristics, and Examples of Anomalies Within Industry}
\label{RQ1}
Even today, the original definition of "anomaly" from the \textit{IEEE Standard 1012 of 1990} \cite{ieeeDefAnomaly1990} persists in active standards such as the \textit{ISO/IEC/IEEE 24765:2017 Systems and Software Engineering -- Vocabulary}. It reads as follows: 

\begin{center}\textit{\fontsize{11pt}{11pt}\selectfont
"Anything observed in the documentation or operation of the software that deviates from expectations \\ based on previously verified software products or reference documents."}\cite{ieeeDefAnomaly1990} 
\end{center}

Some participants [\ref{G}, \ref{H}, \ref{K}, \ref{L}] \textit{argued that it is difficult to define "anomaly"} for them as the definition above does. Nevertheless, our research identified commonalities among the 15 interview participants and showed their resemblances. 

Similarly to the IEEE definition, the aspect of \textit{deviations from expectations} was an essential characteristic of anomalies, also mentioned by several participants [\ref{B}, \ref{C}, \ref{D}, \ref{G}, \ref{J}, \ref{K}, \ref{M}]. This term was also expressed in a similar manner as abnormal behavior [\ref{F}] or described as behavior that is not desired [\ref{I}]. 

An alternative perspective was a mathematical viewpoint, characterizing abnormal data as \textit{outliers from specific trends} [\ref{E}, \ref{M}, \ref{H}, \ref{J}, \ref{M}, \ref{N}]. This was primarily described as data points outside general patterns, such as unusual extreme values, exceeded thresholds, or abnormal decreases/increases. Participant [\ref{G}] mentioned a company-internal standard definition based on specific metrics, including latency, request rates, error counts, system availability, and response times. 

A recurring observation from the interviews was the depiction of anomalies using \textit{synonyms}, particularly failure [\ref{B}, \ref{C}, \ref{D}, \ref{K}, \ref{F}] or error [\ref{C}, \ref{J}, \ref{M}, \ref{N}]. In 2010, the standardization committee \cite{ieeeAnomalyStd2010} classified software anomalies, acknowledging that the original term’s broad meaning led to imprecision and impeded effective communication. Despite concerted efforts to establish precise definitions, nearly every industry interview revealed that these terms were frequently used in an inconsistent manner and often used interchangeably.

Another interesting observation by two participants was that they classified abnormal behavior as an anomaly when it resulted in a \textit{negative effect}. Participant [\ref{B}] defined it as any catastrophic failure. In a similar context, participant [\ref{D}] linked it to disruptions in organizational operations. For instance, decreased user satisfaction, blocked processes, or similar repercussions that needed fast actions [\ref{A},\ref{B},\ref{C}]. Participant [\ref{A}] noted that when something occurred more frequently or happened in critical situations, they placed greater emphasis on resolving it. 

\begin{table}[htbp]
    \columTextSize
\centering  
    \begin{tabular}{@{}m{87mm}@{}} 
    \hline
    \rowcolor{fillColor} \multicolumn{1}{c}{\textbf{Examples}} \\
    \hline
        \labelExamples{B-1}{B-1} - Utilizing a NetApp storage system with \ac{NFS}, the company established shared storage for all \ac{VM}s that ran their services. One routine involved archiving all software versions. This particular action led to an increase in the number of inodes over time, resulting in higher response times across all services.\\
    \hline 
        \labelExamples{C-1}{C-1} - A memory leak due to improper garbage collection occurred during a software release. After some days in production, the application experienced an unexpected failure. This issue remained undetected until an analysis exposed a continuous increase in memory. \\
     \hline
         \labelExamples{D-1}{D-1} -  For some microservices they had self-defined rules, e.g. \ac{RAM}. If a microservice reached a specific threshold, the service was killed and restarted, executed automatically by Kubernetes. \\
     \hline
        \labelExamples{F-1}{F-1} - During an orchestrating process executed on a \ac{VM}, the system unexpectedly exhausted its heap space. The system continued to function without any immediate errors or signs of failure. The only noticeable impact was a substantial decrease in response speed. \\
    \hline
        \labelExamples{F-2}{F-2} -  In a client-server interaction, the user requested some calculation handing over an input entry. The number entered was so large (probably because a key was pressed for too long) that it caused the system to crash. A log with Japanese text indicated a memory overflow error.  \\
    \hline
        \labelExamples{J-1}{J-1}  -  In a range of scenarios, an observable aggregation of diverse jobs within the processing queue was observed. This accumulation had manifested in extended processing times, similar to increased request times. For instance, JS-backend libraries were often the root cause.\\ 
    \hline
        \labelExamples{L-1}{L-1}  - Logging was used to trace abnormal states for comprehensibility and traceability. After implementing a new logging mechanism for anomaly detection, the system failed due to excessive I/O operations. \\
    \hline 
        \labelExamples{M-1}{M-1}  Some applications in their company suffered from poor software elasticity. For instance, the system supported only 500 users at a time. If the $501^{st}$ user tried to log in, the system became slow, possibly unstable, and prone to errors. Eventually, it could crash. \\
    \hline
        \labelExamples{O-1}{O-1}  A frequently executed routine on a CPU that accounted for only 2\% of operations led to excessive use of flash memory, hitting hardware limits. This decreased I/O performance after 10,000 writing cycles led to long-term performance degradation.\\
    \hline 
    \end{tabular}
    \vspace{0.05cm}
\caption{Anomaly examples stated by industrial participants.}
\label{tab:anomalyExamples}
\vspace{-0.8cm}
\end{table}

As outlined in Table \ref{tab:anomalyExamples}, the wide variety of underlying reasons or root causes responsible for the observed anomaly make it difficult to identify anomalies. For instance, the root causes could differ, but the final observed anomaly was similar, such as in the anomaly examples [\ref{L-1}] and [\ref{O-1}], indicating a deviation in the performance of I/O operations. The volatility complicated the development of an accurate and universal anomaly classification that should have helped in predicting \textit{unforeseeable or not reproducible} [\ref{B}, \ref{C}, \ref{K}, \ref{L}, \ref{M}, \ref{O}] anomalies.

\begin{mycolorbox}{Takeaways RQ1: Anomalies in Industry}
Most interview participants described an anomaly as a deviation from expectations where fast resolution was essential to minimize negative effects, which complies with IEEE definitions. Identified examples in Table \ref{tab:anomalyExamples} demonstrate an extract of the variety of often unforeseeable anomalies experienced by interview participants.
\end{mycolorbox}

\subsection*{RQ2: Anomaly Detection Approaches in Industrial Settings}
\label{RQ2}
We distinguish two main approaches to identify anomalies and their root causes used by the interviewees (see Table \ref{tab:interviewees}) and mentioned by industry papers. Anomalies detected by 

\begin{itemize}
    \item \textbf{\textit{rule-based approaches}} rely on thresholds derived from extensive domain knowledge and company-specific insights (e.g., a range of 10\% to 30\% [\ref{D}]), pattern matches (e.g., within a specific time frame [\ref{F}]), or statistical principles (e.g., quantile regression [\ref{K}]), while
    \item \textbf{\textit{\ac{AI}-based approaches}} use supervised and unsupervised AI models \cite{Soldani2022} to detect patterns to identify deviations.
\end{itemize}
Every participant, spanning various company sizes, stated to use self-defined \textit{rule-based} approaches. On the contrary, adopting self-developed \textit{AI-based} approaches was done only by two companies [$1^{st}$ company: \ref{G}, \ref{J}, $2^{nd}$ company: \ref{H}, \ref{K}], even though in the last three years, 80\% of identified industry papers (e.g., \cite{Lee2023,Zhang2023,Catillo2022,Xie2023}) successfully applied various self-developed AI-based algorithms. Four companies, however, did not develop their own AI models but relied on commercially available AI algorithms, e.g., \textit{Dynatrace} used by [\ref{B}, \ref{C}, \ref{E}, \ref{M}, \ref{N}]. One of the drawbacks mentioned with \textit{Dynatrace} was the number of false positives [\ref{B}, \ref{C}, \ref{E}, \ref{M}, \ref{N}]. Participants [\ref{B}, \ref{M}, \ref{N}] received emails when a potential anomaly was detected, but false positive alerts caused them to ignore these emails.

Half of the participants [\ref{B}, \ref{C}, \ref{D}, \ref{M}, \ref{N}, \ref{J}, \ref{O}] mentioned that they preferred rule-based approaches due to the widespread availability of established monitoring tools that gather essential monitoring parameters and visualize them. Furthermore, they could define rules for these parameters that trigger alerts. The participants currently opted for this approach because they required less computational costs than \ac{AI} for training [\ref{J}] (which required retraining to adapt to the latest input [\ref{G}]) and detected anomalies in nearly real-time [\ref{D}, \ref{F}, \ref{K}]. 

On the contrary, eight participants [\ref{C}, \ref{D}, \ref{F}, \ref{H}, \ref{J}, \ref{M}, \ref{N}, \ref{O}] stated that their rule-based approaches required extensive domain knowledge and expertise to define valuable rules and thresholds. Thus, setting up these rules was time-consuming, based on subjective insights, and might not detect unforeseen anomalies. For instance, participant [\ref{H}] stated that one of their rules detected \texttt{sudo} operations. However, the rule did not specify \texttt{su} where they missed these anomalies. Participant [\ref{G}] experimented at his company with a self-implemented \ac{AI} approach to identify and detect irregular patterns. Using K-means mainly allowed the company to reasonably detect anomalies in that scope, but their approach was not deemed production-ready due to shortcomings in their training dataset. Thus, to effectively use AI-based approaches, it was crucial to enhance dataset quality by understanding input parameters [\ref{G}], allowing to maximize AI approaches' potential [\ref{F}]. Additionally, participant [\ref{O}] suggested that AI could optimize the dataset size by finding relevant monitoring parameters that did not introduce bias.

\begin{mycolorbox}{Takeaways RQ2: Anomaly Detection Approaches}
While research mainly covered new implementations of AI algorithms for anomaly detection, companies used particularly rule-based or commercially available AI-based approaches. The main reasons for the interviewed companies using rule-based approaches were that they had lower computational costs, wider tool support, high adaptability to domain constraints through self-defined rules, and fast anomaly exposure. These rules, however, were tailored to each company and hinged on a deep understanding of the parameters and thresholds involved. Commercial AI-based approaches could offer more flexible solutions to identify patterns and anomalies in various contexts. Nonetheless, the dataset's quality limited the effectiveness of both rule-based and AI-based approaches.
\end{mycolorbox}

\subsection*{RQ3: Runtime Monitoring Data for Anomaly Detection}
\label{RQ3}
As identified in \hyperref[RQ2]{RQ2}, the dataset's quality is the foundation of a good anomaly detection approach where using key monitoring parameters and understanding their relationship helped to avoid false positives. For instance, participant [\ref{B}] illustrated that increased user requests resulted in slower response times. Yet, it should not have been considered an anomaly as the relationship between these parameters remained unchanged. Therefore, this section elaborates on key monitoring parameters extracted from the systems' runtime data to identify anomalies. We categorized these parameters into \textbf{logs, traces,} and \textbf{metrics}. \textcolor{logs}{\textbf{Logs}} are predefined semi-structured emitted messages that contain a specific timestamp, verbosity level (such as \texttt{INFO, WARN, DEBUG}, and \texttt{ERROR}), and unstructured natural language, including comprehensive information about an event specified by developers [\ref{C}, \ref{F}, \ref{H}] \cite{Zhang2023RobustFailureDiagnosis}. A \textcolor{traces}{\textbf{trace}} is a microservice-specific data type that represents the end-to-end execution of a single request, traversing through various microservices. A span is a trace segment encompassing more specific metadata (e.g., start and end time). \textcolor{metrics}{\textbf{Metrics}} consist of numeric time series data of system instances by collecting various performance data via, for instance, Prometheus, \cite{Zhang2023RobustFailureDiagnosis} [\ref{C}, \ref{D}, \ref{E}, \ref{G}, \ref{H}, \ref{J}, \ref{L}]. These three primary monitoring data types could be measured via third-party software. For instance, participants rely on Heroku, Dynatrace, OpenShift, Nagios, Grafana, Kubernetes, Prometheus, Jaeger, GrayLog to collect, track and visualize their runtime monitoring data.

Regardless of the specific third-party software or monitoring data type, various parameters could be calculated from different monitoring data types. For instance, response time can be identified or calculated via logs, traces, or metrics [\ref{A}, \ref{C}]. Different errors were visible in logs via their verbosity level and in traces via their HTTP status [\ref{J}]. Out-of-memory issues could be identified via logs or metrics \cite{Huang2023}. 

\begin{figure}[htp]
    \centering
    \hspace*{-0.6cm}\includegraphics[width=1.15\columnwidth]{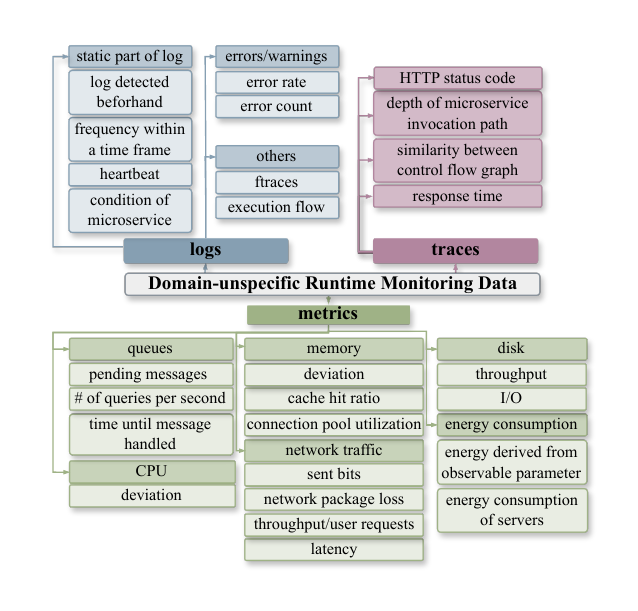}
    \vspace{-1.3cm}
    \caption{Runtime monitoring data types categorization into logs, traces, metrics, and their associated parameters.}
\label{fig:monitoringDataCategorisation}
\vspace{-0.6cm}
\end{figure}

We depict the three monitoring data types and their associated key monitoring parameters for anomaly detection in Figure \ref{fig:monitoringDataCategorisation}, further elaborated in the following subsections.

\subsubsection{Logs}
Half of the participants looked at logs to identify anomalous behavior to troubleshoot their applications, and one ([\ref{F}]) solely relied on logs. 

Due to the semi-structured characteristics of logs and the extensive volume [\ref{L}] \cite{Catillo2022,Cinque2022}, four participants [\ref{F}, \ref{H}, \ref{J}, \ref{O}] extracted the \textcolor{logs}{\textbf{static part of logs}} for further processing. The participants then either identified if this log was detected beforehand (true/false) [\ref{F}] or looked at the frequency within a time frame [\ref{H}, \ref{O}] \cite{Jia2017,Jia2017-LogSed,Shan2019,Cinque2022,Catillo2022}. For the frequency, participants [\ref{F}, \ref{J}, \ref{L}] counted the number of occurrences of a static part of the log. Participant [\ref{F}] mentioned two ways to count it, either via event-based or sequence-based approaches. In his company, they relied on event-based counts where the static log got counted when detected, which was feasible for their single-threaded processes where logs occurred in their sequential order. Participant [\ref{L}] chose a sequence-based approach because they had a multi-threaded application where the order of the logs could change. When only counting the occurrences as done with the event-based approach and without considering the time (fixed or sliding time frame), deadlocks, or the anomaly example [\ref{M-1}], that dynamic scaling of the thread pool slowed down response times in Open Liberty could not get detected. In addition, thread names or their ID could be included in the log line [\ref{L}].

However, when calculating the frequency, participant [\ref{F}] mentioned that some parts of the software could not be used as regularly as others, resulting in different numbers of counts, where zero occurrences were also feasible. Participant [\ref{F}] recorded regular heartbeats to counteract overseeing an involuntary count of zero logs. Participant [\ref{O}] further indicated that looking at the condition (e.g., running, idle) of a microservice via logs over time allowed him to identify if the anticipated conditions occurred as expected.

Furthermore, participants relied on information regarding \textcolor{logs}{\textbf{errors or warnings}} extracted from logs. According to participants [\ref{E}, \ref{G}, \ref{J}], error-specific parameters included error rate (calculated as the number of errors divided by the total number of occurred verbosity levels indicated in percentage) and the counted number of errors within a specific time frame [\ref{G}, \ref{O}]. Participant [\ref{J}] did not differentiate the type of error when counting the occurrences. However, participant [\ref{F}] stated that they analyzed this parameter in the log message because the term \textit{error} did not necessarily signify a system deviation (e.g., error handling routine started) as indicated in the anomaly description [\ref{F-2}]. This example also showed that the absence of the English term \textit{error} in the logs message did not indicate the absence of it.

\textcolor{logs}{\textbf{Other}} potential approaches involved extracting ftraces to understand kernel operations, especially since Participant [\ref{I}] utilized \acf{PLC} as their programming language. Furthermore, logs could also capture the execution flow of inter and intra services similarly to traces where task or transaction IDs tied logs together \cite{Jia2017,Jia2017-LogSed}. \customnewline

\subsubsection{Traces}
Participants [\ref{D}, \ref{G}, \ref{J}, \ref{M}, \ref{N}] monitored microservice architectures and, therefore, could also rely on traces. 

Similar to the error parameter extracted from logs, \textcolor{traces}{HTTP status code} allowed interview participants [\ref{H}, \ref{J}, \ref{N}] and \cite{Li2021,Zhang2023RobustFailureDiagnosis} to extract the frequency of errors from endpoints, such as 500 (internal server error) or the availability of services \cite{Ma2020}, such as 2xx (successful), or 4xx (client error). 

For calculating further parameters, a control flow graph based on information regarding traces, microservices, and their relation to each other was recreated [\ref{D}, \ref{J}], where literature often specified the graph as a directed acyclic graph \cite{Liu2020,Nedelkoski2019a,Jin2020,Meng2021,Wang2020,Zhang2023,Xie2023a}. Participant [\ref{N}] stated that generating the graph manually required extensive system knowledge.

Based on this graph, parameters such as the \textcolor{traces}{\textbf{depth of the microservice invocation path}} could be calculated [\ref{D}, \ref{J}]. Participant [\ref{J}] stated that this parameter was essential when diving deeper into the root cause of an anomaly. For instance, more spans than anticipated could indicate an anomaly. 

Industry papers revealed further parameters that were not identified during the interviews. For instance, one parameter summarised the \textcolor{traces}{\textbf{similarity between the control flow graph}} of the microservices invocation pattern, such as if the in and outgoing dependencies remained the same for similar requests or microservices were missing within a trace \cite{Xie2023,Ding2023,Bento2021}. 

Furthermore, the \textcolor{traces}{\textbf{response time}} calculated via traces was an indicator for anomalous behavior \cite{Nedelkoski2019,Nedelkoski2019a,Xie2023,Liu2020}. For instance, several industry papers looked at the deviation of the response time when finding a matching invocation path \cite{Jia2017,Jia2017-LogSed,Liu2020,Nedelkoski2019a,Jin2020,Zhang2023}. Response times could be measured for the processing time of each microservice \cite{Jin2020,Zhang2023} or for the whole request (calculated by the time difference between the incoming request and the application's response) [\ref{E}, \ref{N}]. The latter was not exclusive to a microservice architecture but could also be calculated for a monolith. For instance, logs could contain information regarding a request and associated response where the time between these logs was calculated [\ref{N}]. Participant [\ref{K}] also collected idle times within the response times and latency \cite{Xie2023a}.
As in the anomaly example [\ref{F-1}] with the exhausted heap space or example [\ref{B-1}] with accumulated inodes, deviations in system-wide response time could be an early indicator of an accumulating anomaly and could prevent system failures as in the anomaly example [\ref{M-1}]. If the response time increased, the load for one service could be too high [\ref{A}, \ref{C}]. The response times of external systems, such as a database, were of further interest [\ref{E}] \cite{Ma2020}. \customnewline

\subsubsection{Metrics}
All participants mentioned several metric parameters. We assume metrics were mentioned often because these were already available, intuitive, monitored over time with a fixed interval, and did not require extensive calculations. Furthermore, we assume that because all participants used rules and thresholds, they had more experience with metrics. 

Several participants indicated that they observed different parameters regarding their \textcolor{metrics}{\textbf{queues}} [\ref{F}, \ref{J}, \ref{L}]. For instance, in the anomaly example [\ref{J-1}], a processing queue aggregated too many jobs, indicating an anomaly. Participant [\ref{F}] analyzed how many pending messages were in the queue and looked at the number of resolved queries per second within the queue \cite{Liu2021}. Participants [\ref{A}, \ref{E}] focused on the time it took until the server handled the message in the queue (queue time).\customnewline

In terms of \textcolor{metrics}{CPU}, a deviation or unusual pattern thereof could signal an anomaly [\ref{C}, \ref{F}, \ref{H}] \cite{Lee2023,Li2021}. Participant [\ref{C}] observed that a sudden or gradual increase could indicate a potential anomaly. However, participant [\ref{H}] stated that a decrease in CPU utilization could also indicate an anomaly, given that their CPUs were typically nearly 100\% utilized in normal production. The CPU-related parameter could further be split into CPU user usage, CPU system usage, CPU wait, or CPU throttling [\ref{H}, \ref{K}] \cite{Lee2023}. However, participant [\ref{A}] disregarded CPU usage because this only indicated necessary resource scaling. 

When referring to \textcolor{metrics}{\textbf{memory}}, participants [\ref{C}, \ref{D}, \ref{J}] and \cite{Lee2023} stated that they also looked at deviations of memory and memory dump. In the anomaly example [\ref{C-1}], a memory leak from improper garbage collection was detected by monitoring memory consumption. Participant [\ref{D}] monitored their sensor memory so that they did not allocate more resources during runtime than during testing. Kubernetes took over the resource allocation via soft and hard thresholds where services could get restarted several times [\ref{D-1}]. In addition, [\ref{H}] indicated that the cache hit ratio, in combination with response time, revealed anomalies. Participants [\ref{E}, \ref{H}, \ref{N}] stated that they looked at connection pool utilization and if there was an unusual amount of these connections, for instance, to a database service.\customnewline

Participant [\ref{H}] mentioned \textcolor{metrics}{\textbf{network traffic}}, where a change in the transmitted bits or network package loss could indicate an anomaly. Lee et al. \cite{Lee2023} emphasized using network throughput, and two participants focused on increases in network traffic from increased user requests or timeout messages [\ref{J}, \ref{L}]. Participant [\ref{E}] further identified latency regarding the network or network throughput \cite{Lee2023,Li2021}. \customnewline 

\textcolor{metrics}{\textbf{Disk}} parameters were also a good indication for identifying a deviating behavior, as participants [\ref{F}, \ref{H}] indicated. For instance, [\ref{H}] looked at disk throughput and disk I/O, such as I/O wait, idle, and the device read speed \cite{Lee2023}. Participant [\ref{I}] relied on Linux-based metrics measured via \texttt{iotop}, watching I/O usage information output by the Linux kernel to measure stress and increased load on hard disks. \customnewline

During the literature review, we identified that \textcolor{metrics}{\textbf{energy consumption}} was not considered so far for anomaly detection. Thus, we explicitly asked our interview participants if they already employed this parameter in their anomaly detection approach. So far, one participant measured their hardware's energy consumption. However, they stated that they did not try to measure their processors' energy consumption, for instance, with \ac{RAPL} [\ref{I}]. Several participants also did not measure this type of metric. However, they saw this as a potential anomaly indicator and were keen to explore it further [\ref{C}, \ref{D}, \ref{F}, \ref{K}, \ref{L}]. 

However, participants [\ref{G}, \ref{H}, \ref{I}, \ref{J}] argued that cloud-based deployment made it impossible to physically measure energy consumption by themselves nor provided these parameters. For instance, AWS or Google billed based on hourly usage without providing insights into energy consumption. Participant [\ref{D}] also mentioned lacking tools to monitor server energy.

Contrary, Participant [\ref{C}] believed measuring energy could be feasible. Participants [\ref{D}, \ref{M}, \ref{N}] suggested that already observable metrics, like CPU or memory, could assist with calculating this parameter. However, optimization strategies could distort the energy consumption parameter. Thus, a potential anomaly might have occurred when an increase in energy outside of the threshold was monitored.

Participants [\ref{K}, \ref{L}], however, suggested that considering server energy consumption could provide insights into how other systems affected the software under test. However, accurate measurement and interpretation were crucial, as participant [\ref{B}] noted that server and \ac{VM} energy consumption significantly fluctuated. Thus, it was challenging to map energy consumption to anomalies due to many influencing factors (e.g., energy used by unrelated monitoring tools on the same server) [\ref{B}, \ref{H}]. In addition, participant [\ref{C}] noted modern solutions targeting energy reduction shut down unused applications and restarted them upon demand. Consequently, zero energy usage should not have been considered an anomaly in such scenarios. Measuring energy consumption in embedded systems was intricate due to minuscule units (e.g., Milliamperes) where temperature could affect consumption [\ref{O}]. 

\subsubsection{General Remarks}
We recognized that the mentioned parameters did not differ based on the participants' domain, used approach, or whether they monitored a microservice or monolithic architecture [\ref{B}, \ref{D}, \ref{H}, \ref{M}, \ref{N}]. A monolith could be seen as one microservice where logs or metrics, for instance, cannot be assigned to one specific service but were measured for the whole monolith (e.g., when deployed via a container [\ref{B}]). Also, microservice-based systems could first look at system-wide parameters, diving deeper into more specific information for each microservice [\ref{D}]. However, it was essential to note that orchestration tools, such as Kubernetes, could influence the system behavior and, thus, ultimately, the observed parameters due to different scheduling of resources or automated scaling [\ref{H}]. Thus, considering parameters collectively and their correlation was essential to avoid falsely identifying desired behavior as an anomaly. 

Moreover, as discussed by participants [\ref{H}, \ref{I}, \ref{K}], an optimal sampling interval was crucial, avoiding excessive fluctuation when monitoring runtime data too frequently. Participant [\ref{H}] and \cite{Ma2020} noted that a wide sampling interval might have missed deviations, and participant [\ref{K}] indicated that the interval had to be chosen sensibly to prevent excessive data collection. Participant [\ref{M}] added that the aggregation of parameters over a time frame (e.g., average, etc.) also influenced how well anomalies were displayed.

\begin{mycolorbox}{Takeaways RQ3: Key Monitoring Parameters}
 Runtime monitoring data consists of three data types - logs, traces, and metrics. Figure \ref{fig:monitoringDataCategorisation} illustrates the key monitoring parameters to identify anomalies.
\end{mycolorbox}

\section{Threats to Validity}
\label{sec:threatsToValidity}
This section discusses the Threats to Validity according to Wohlin et al. \cite{Wohlin2012} and illustrates how we mitigated them. 

We avoid a lack of \textbf{Internal Validity} using data triangulation via a literature study and interviews. Therefore, we extended the literature study by \cite{Soldani2022} and included 36 industry papers. To mitigate the papers` selection bias, a second author conducted a blind review as indicated in Section \ref{sec:systematicReview}. Furthermore, we gained insights into 12 companies via 15 interview participants, where we mitigated coding bias by discussing the generated codes and assignment thereof with two authors. However, the sample size consists of 15 participants, which might limit the generalizability of the findings. However, this study aimed to gain in-depth insights and understand the participants' experiences and perspectives. The focus was on exploring detailed data rather than achieving statistical generalizability. We considered that this sample size could emphasize atypical responses, which we mitigated by providing links to the participants who made this statement. Furthermore, four interview participants have under five years of experience in this area, which could seem to render them less suitable. However, not only did the contacted company representative explicitly forward us these contacts, but also participants [\ref{G}, \ref{H}, \ref{K}] focused on research regarding enhancing their companies' anomaly detection approach.

We enhanced the \textbf{External Validity} by selecting interview participants from different domains (Finance, Web, Hardware, Others), software architectures, company sizes, and experiences. Although participants mentioned different root causes of anomalies, especially in the hardware domain, the observed anomalies and parameters mirrored those in other domains, revealing no significant differences in anomaly detection strategies or parameters. In addition, the case studies identified in the literature further enhanced the applicability of anomaly detection approaches and parameters in various domains.

To minimize the threat to \textbf{Construct Validity}, we identified that numerous parameters could be derived from various runtime monitoring data types depending on the interview participants' monitoring strategies, complicating their classification. Thus, we provided detailed parameter descriptions with additional input when the parameter could get extracted from other monitoring data types and calculation methods to address this (e.g., response time calculated via logs or traces, errors identified via logs and traces).

Regarding \textbf{Conclusion Validity}, the obtained key runtime parameters were more difficult to identify for \ac{AI} algorithms presented in published literature because the authors nearly always did not provide a dataset, code to the algorithm, or discuss their chosen input. Furthermore, \ac{AI}-based approaches did not necessarily require dissecting the dataset into single parameters but entrust the \ac{AI} model to make sense of the monitoring data. In addition, interview participants defined rules on their parameters, providing more knowledge of key monitoring parameters for their anomaly detection approach.

\section{Conclusion}
\label{sec:conclusion}
Software systems must function reliably in industry, although anomalies are becoming more prevalent. Analyzing runtime monitoring data, including logs, traces for microservices, and metrics, is challenging due to the required excessive knowledge of the system and the huge amount of collected monitoring data. Detecting these unpredictable anomalies as soon as possible avoids more serious system failures. Therefore, it's crucial to understand anomalies in various domains, how they are detected in the industry, and which key monitoring parameters extracted from runtime monitoring data depict anomalies to enhance anomaly detection methods.

Therefore, we extended a literature survey, resulting in 36 relevant industry papers, and executed 15 semi-structured interviews across various domains. \hyperref[RQ1]{RQ1} identified that our industrial interview participants described an anomaly as an unforeseeable deviation from expectations that has outliers of specific trends in monitoring data and has a negative effect. The interview participants provided several anomaly examples. Regarding \hyperref[RQ2]{RQ2}, our analysis revealed that all 12 companies applied rule-based anomaly detection, whereas additionally two companies developed an internal AI-based approach, and four companies relied on commercial AI-based approaches. Notably, the existing literature predominantly focused on AI-based approaches within the last three years. For all approaches, a deep understanding of the key monitoring parameters and thresholds is essential to improve both anomaly detection approaches. Thus, based on \hyperref[RQ3]{RQ3}, we provide a concise summary of key monitoring parameters (collected in Figure \ref{fig:monitoringDataCategorisation}) extracted from runtime monitoring data types (logs, traces, and metrics).\\

As identified, it is essential to understand and improve the quality of the collected datasets by understanding key monitoring parameters, their relationships, and their influences on the system or microservice. Therefore, \textbf{future work} will establish a statistical model based on these parameters that provides insights into the parameters' relationship and necessity. Furthermore, cause-effect relations of an anomaly and the resulting system behavior should be calculated. To validate the explainability model, we will employ a benchmark system, TrainTicket \cite{Zhou2021}, with injected anomalies in a controlled environment and a case study of real-world data comprising logs, traces (Jaeger), and metrics (Prometheus). Given energy consumption's potential as a new anomaly-detection parameter, we will consider \textit{\ac{RAPL}} and try to measure Watt with an external voltage metering tool. Based on the new insights, we can make informed decisions on their dataset and propose additional rules for detecting anomalies in the company.

\bibliographystyle{IEEEtran} 
\linespread{0.814}
\bibliography{references}
\end{document}